\documentclass[
 reprint,            
 superscriptaddress, 
 amsmath,amssymb,    
 aps,                
 prl,                
 floatfix            
 ,nofootinbib
]{revtex4-2}

\usepackage{microtype}  
\usepackage{color, verbatim}

\usepackage{latexsym}
\usepackage{amsmath}
\usepackage{amssymb}
\usepackage{bm}
\usepackage{adjustbox}
\usepackage{multirow}
\usepackage{mathrsfs}
\usepackage{enumerate}
\usepackage{slashed}
\usepackage{graphicx}    
\usepackage{subcaption}    
\usepackage[font=small,labelfont=bf,justification=centerlast]{caption}
\usepackage{wrapfig}     
\usepackage[normalem]{ulem}

\usepackage{tikz}
\usepackage[compat=1.1.0]{tikz-feynman}

\def\bea{\begin{eqnarray}}
\def\eea{\end{eqnarray}}

\newcommand{\nn}{\nonumber}

\def\ie{\textit{i.e.}}

\usepackage[normalem]{ulem}

\RequirePackage[colorlinks=true,urlcolor=blue,anchorcolor=blue, citecolor=blue,filecolor=blue, linkcolor=blue, menucolor=blue, linktocpage=true, pdfproducer=medialab, pagebackref=true]{hyperref}
\hypersetup{
  colorlinks   = true, 
  urlcolor     = mygreen, 
  linkcolor    = blue, 
  citecolor   = blue 
}
\usepackage{orcidlink}

\usepackage[capitalise]{cleveref}
\crefname{table}{Table}{Tables}
\crefname{equation}{Eq.}{Eqs.}
\crefname{appendix}{App.}{Apps.}
\crefname{section}{Sec.}{Secs.}
\crefname{figure}{Fig.}{Figs.}

\usepackage{csquotes}
\definecolor{mygreen}{rgb}{0.0, 0.5, 0.0}

\definecolor{mygreen}{rgb}{0.0, 0.5, 0.0}

\newcommand{\magenta}[1]{\color{magenta} #1 \color{black}}

\usepackage[font=small,labelfont=bf,justification=centerlast,singlelinecheck=true]{caption}
\usepackage{mathtools}

\usepackage{cleveref}
\crefname{table}{Table}{Tables}
\crefname{equation}{Eq.}{Eqs.}
\crefname{appendix}{App.}{Apps.}
\crefname{section}{Sec.}{Secs.}
\crefname{figure}{Fig.}{Figs.}
\usepackage[normalem]{ulem}
\usepackage{physics}
\usepackage{slashed}
\usepackage{gensymb}
\usepackage{wasysym}
\usepackage{bm}
\usepackage{dsfont}
\usepackage{pifont}
\usepackage[most]{tcolorbox}
\definecolor{light-gray}{gray}{0.9}

\begin{document}
\vspace*{-1cm}
\begin{flushleft}
\magenta{IFT-UAM/CSIC-25-102}
\end{flushleft}

\title{Taming forward scattering singularities in partial waves}

\author{Marta Fuentes Zamoro\orcidlink{0009-0001-3155-8205}}
\email{marta.zamoro@uam.es}
\affiliation{Departamento de F\'isica Te\'orica and Instituto de F\'isica Te\'orica UAM/CSIC, Universidad Aut\'onoma de Madrid, Cantoblanco, 28049, Madrid, Spain}

\author{Benjamín Grinstein\orcidlink{0000-0003-2447-4756}}
\email{bgrinstein@ucsd.edu}
\affiliation{Department of Physics, University of California, San Diego, La Jolla, CA 92093, USA}
\author{Pablo Quílez\orcidlink{0000-0002-4327-2706}}
\email{pquilezlasanta@ucsd.edu}
\affiliation{Department of Physics, University of California, San Diego, La Jolla, CA 92093, USA}


\begin{abstract}
   Perturbative partial-wave amplitudes diverge in cases with a massless exchanged particle in the $t$-channel. We argue that the divergence is an artifact of perturbation theory and give a prescription for the all-orders correction factor that renders the partial waves finite. As an example, we apply this to longitudinal $W^+W^-$ elastic scattering, for which there is a photon exchange $t$-channel contribution, and derive improved \emph{quasi-perturbative} unitarity bounds on the mass of the Higgs. The method is also useful for $t$-channel exchange of very light particles.
\end{abstract}

\maketitle

The partial-wave decomposition plays an important role in many advances in particle physics. It is used in analysis of experimental data to uncover the quantum numbers of resonances seen in scattering. On the theory side, it is used extensively in exploiting the consequences of unitarity on scattering amplitudes. In this vein, the conditions that unitarity imposes on partial waves were central to establishing upper bounds on scattering amplitudes at asymptotically large center of mass (CM) energies  \cite{Froissart:1961ux,Martin:1969ina}. Famously, partial-wave unitarity was used to establish the strength of Higgs self interactions, or, equivalently, the Higgs mass, for which perturbation theory remains applicable \cite{Lee:1977eg,Lee:1977yc,Dicus:1973gbw}. Among many other examples, an extension of that idea has been more recently used to obtain bounds on di-boson scalar resonances \cite{DiLuzio:2017qgm}, on multi-Higgs extensions of the Standard Model (SM) \cite{Grinstein:2013fia,Grinstein:2015rtl}, on the strength of ALP interactions \cite{Brivio:2021fog} and on dimension-6 and -8 operators in an effective field theory extension of the SM (SMEFT) \cite{Corbett:2014ora,Corbett:2017qgl,Remmen:2019cyz,Cohen:2021gdw}. 

Unitarity bounds at the electroweak scale play a key role in defining the Higgs Effective Field Theory (HEFT) which displays non-decoupling new physics \cite{Cohen:2021ucp,Banta:2021dek}, in contrast to SMEFT in which the new physics decouples.

However, when a massless particle is exchanged in the $t$-channel, the resulting scattering amplitude displays a singularity in the forward direction, a non-integrable divergence that renders the partial-wave amplitudes ill-defined. This is of course well known: it is precisely the same divergence displayed by the Rutherford scattering amplitude. In this letter we give a prescription that renders the partial waves finite. The clue to the resolution is to include in the computation non-perturbative effects. Short of proving this assertion, we will present evidence that we think is both suggestive and persuasive, and hope to come back to providing a proof in the future. 

The partial-wave amplitude for elastic scattering is given by 
\begin{equation}
    a_{\ell}=\frac{1}{32\pi} \int_{-1}^1 \!\!{\rm d}\!\cos{\theta}\; P_{\ell}(\cos\theta) \,\mathcal{T}(\sqrt{s},\cos{\theta})\,.
    \label{eq:pwa} 
\end{equation}
For every pair of identical particles in either the initial or final state, an additional factor $1/\sqrt{2}$ must be added. The unitarity condition for elastic scattering is 
\begin{equation}
\label{eq:uni-bound-aJ}
      |a_\ell-i|^2\le1\,,
\end{equation}
and, since tree-level amplitudes are real, this is customarily replaced by 
\begin{equation}
\label{eq:uni-tree-bound-aJ}
(a_{\ell})^2\le1\,.
\end{equation}

The tree-level expression for the partial-wave amplitude in Eq.~\eqref{eq:pwa} exhibits a logarithmic divergence for even $\ell$ when there is a $t$-channel massless particle exchange contribution to the tree-level scattering amplitude, 
 \begin{equation}
 \label{eq:T-divergence}
 \mathcal{T}^{(0)}\propto\frac{1}t\propto \frac{1}{1-\cos\theta}\,.
 \end{equation}

We propose solving this problem by replacing the tree-level amplitude $\mathcal{T}^{(0)}$ for a photon exchange between particles of charge $z_1$ and $z_2$ as follows: 
 \begin{equation}
     \label{eq:T-corrected}
 \mathcal{T}^{(0)}\to \mathcal{T}^{(0)}e^{iW}\,,
 \end{equation}
 where\footnote{We have used the sign of the phase in \cite{messiah2014quantum}, which is opposite to that of \cite{gottfried2003quantum}. Our choice is consistent with those of Dalitz, \cref{eq:DalitzPhase}, and Weinberg, \cref{eq:Weinberg}.}  
 \begin{equation}
    \label{eq:phase-coef-W}
  i W= \Big(\epsilon-i\frac{z_1z_2\alpha_{\rm em}}{\beta}\Big)
   \ln\big[(1-\cos\theta)/2\big]
\end{equation}
and $\beta=\frac{\sqrt{(p_1\cdot p_2)^2-m_1^2m_2^2}}{p_1\cdot p_2}$  is the relative velocity of the colliding (or, equivalently, the final state) particles and $\epsilon\to0^+$. Although the divergent denominator is still present, the phase factor $\exp{iW}$ is an oscillating function that tames the divergence, making it integrable. Note that if regulated by a small mass $m$ in the propagator of the particle exchanged in the $t$-channel, the resulting unitarity bounds depend sensitively on the value of the mass $m$ and become meaningless in the limit $m\to 0$~\cite{Blas:2020dyg}.

To justify this conjecture, consider first the classic example of non-relativistic Coulomb potential scattering. As is well-known, the Born scattering amplitude $f_C^{(0)}(k;\theta)$ exhibits the forward divergence in \cref{eq:T-divergence}.  However, the exact solution to the problem is known \cite{messiah2014quantum,gottfried2003quantum,schiff1955quantum,rose1961relativistic},\\
 \begin{equation}
  \label{eq:fC}
  f_C(k;\theta)=\gamma \frac{\Gamma(1-i\gamma)}{\Gamma(1+i\gamma)}
  \frac{e^{-i\gamma\ln[(1-\cos\theta)/2]}}{k(1-\cos\theta)}\,,
\end{equation}
 where $\gamma=z_1z_2\alpha_{\rm em}/v$ and the velocity $v=k/m$ is determined from the asymptotic non-relativistic kinetic energy,
$E=k^2/2m=\tfrac12mv^2$. 
We note that
\begin{equation}
f_C(k;\theta)= \Bigg[\frac{\Gamma(1-i\gamma)}{\Gamma(1+i\gamma)}
  e^{-i\gamma\ln[(1-\cos\theta)/2]} \Bigg]\, f_C^{(0)}(k;\theta)\,.
  \label{Eq:SchrodingerPhase}
\end{equation}
The pre-factor in square brackets includes the phase $\exp{iW}$ times a phase  $\Gamma(1-i\gamma)/\Gamma(1+i\gamma)$ that can be expanded in powers of the coupling to give a correction of higher order in $\alpha_{\rm em}$.  For this problem, the Coulomb phase shift, $\delta_l$, is given by \[
e^{i\delta_l} =\frac{\Gamma(l+1-i\gamma)}{\Gamma(l+1+i\gamma)}\,.
    \]
The textbook derivation of this result involves the determination of the expansion in spherical harmonics of the exact solution of the Schr\"odinger equation \cite{messiah2014quantum,gottfried2003quantum,schiff1955quantum,rose1961relativistic}. We have verified that the same result can be obtained by integrating directly the full scattering amplitude with a shift $i\gamma\to i\gamma+\epsilon$ as in \cref{eq:T-corrected,eq:phase-coef-W}.\footnote{We have not found this method for deriving this result, that uses only the asymptotic form of the wavefunction, elsewhere in the literature.}

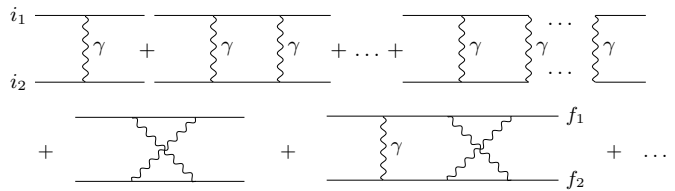
\begin{figure}[t]
    \centering

\resizebox{0.48\textwidth}{!}{
    \begin{tikzpicture}
 \hspace{-15pt} 
 \begin{feynman}
    \diagram*{
      e1 [particle=\(i_1\)] --  a --  e2 [particle],
      e3 [particle=\(i_2\)] --  b --  e4 [particle],
      a -- [photon, edge label=\(\gamma\)] b,
    };
  \end{feynman}
\end{tikzpicture}
\raisebox{1.9em}{\hspace{-30pt} $+$\hspace{-10pt}}
\begin{tikzpicture}
  \begin{feynman}
    \diagram*{
      e1 [particle] --  a  --  a2 --  e2a [particle=\({\color{white}i_1}\)],
      e3 [particle] --  b  --  b2 --  e4a [particle=\({\color{white}i_2}\)],
      a -- [photon, edge label=\(\gamma\)] b,
      a2 -- [photon, edge label=\(\gamma\)] b2,
    };
  \end{feynman}
\end{tikzpicture}
\raisebox{1.9em}{\hspace{-20pt} $+ \, \dots\,  +$\hspace{-10pt}} 
\begin{tikzpicture}
  \begin{feynman}
   \diagram*{
      e1 [particle] --  a -- a0  --[draw=none,edge label'={$\dots$}]  a2 --  e2a [particle=\({\color{white}i_1}\)],
      e3 [particle] --  b -- b0  --[edge label={$\dots$},draw=none]  b2 --  e4a [particle=\({\color{white}i_2}\)],
      a -- [photon, edge label=\(\gamma\)] b,
      a0 -- [photon, edge label=\(\gamma\)] b0,
      a2 -- [photon, edge label=\(\gamma\)] b2,
    };
  \end{feynman}
\end{tikzpicture}}
\resizebox{0.48\textwidth}{!}{\raisebox{1.9em}{ $+$}
\begin{tikzpicture}
  \begin{feynman}
    \diagram*{
      e1 [particle] --  a  --  a2 --  e2a [particle=\({\color{white}i_1}\)],
      e3 [particle] --  b  --  b2 --  e4a [particle=\({\color{white}i_2}\)],
      a -- [photon] b2,
      a2 -- [photon] b,
    };
  \end{feynman}
\end{tikzpicture}
\raisebox{1.9em}{\hspace{-4pt} $+$\hspace{4pt}}
\begin{tikzpicture}
  \begin{feynman}
    \diagram*{
      e1 [particle] --  a0 -- a  --  a2 --  e2a [particle=\(f_1\)],
      e3 [particle] --  b0 -- b  --  b2 --  e4a [particle=\(f_2\)],
      a -- [photon] b2,
      a0 -- [photon, edge label=\(\gamma\)] b0,
      a2 -- [photon] b,
    };
  \end{feynman}
\end{tikzpicture}
\raisebox{1.9em}{ $+$\quad\dots}
}
    \caption{In the non-relativistic limit, the phase  $e^{iW}$, see \cref{Eq:SchrodingerPhase}, arises from the resummation of the ladder diagrams (top line); the crossed diagrams do not contribute in the non-relativistic limit (bottom line).{\hfill}
    }
    \label{fig:LadderAndCrossed}
\end{figure}

In the non-relativistic case, the phase factor $\exp(i W)$ in \cref{eq:phase-coef-W} can be understood as arising from the  resummation of long-range (soft) photon exchanges, represented by the ladder  diagrams in \cref{fig:LadderAndCrossed}. 
Indeed, the first Born approximation $f_C^{(0)}(k;\theta)$ corresponds to a single-photon exchange (first diagram in \cref{fig:LadderAndCrossed}). The uncrossed ladders can be summed via an integral equation, the Bethe--Salpeter equation~\cite{PhysRev.84.1232,PhysRev.84.350}, which reduces to the Schrödinger equation in the non-relativistic limit, connecting ladder resummation to the phase $\exp(i W)$~\cite{Weinberg:1995mt}.

There is some additional evidence that this conjecture works in the relativistic case.  Dalitz computes the relativistic scattering of a charged particle $z_1$ off of
a Coulomb potential due to a nucleus of charge $z_2$ to 1-loop order using the language of quantum
field theory \cite{Dalitz:1951ah}. He regulates the IR via a photon mass $m$. Reverting to rationalized Gaussian units  and retaining only the forward log-divergence,
his result gives, as $m\to0$

{\fontsize{9.53pt}{11.7pt}\selectfont
\begin{equation}
  \mathcal{T}^{(1)}=\frac{z_1z_2 e^2\gamma^0}{m^2+2p^2(1-\cos\theta)}
  \bigg[1-i\frac{z_1z_2\alpha }{p/E}\ln\Big(\frac{2p^2(1-\cos\theta)}{m^2}\Big)\bigg]\,.\nonumber
  \end{equation}}

\noindent This is consistent with our proposal in \cref{eq:T-corrected,eq:phase-coef-W}. Dalitz conjectures that when including higher-order corrections, the result exponentiates, replacing the term in brackets by 
\begin{equation}
\label{eq:DalitzResum}
   \mathcal{T}^{(1)}\to \mathcal{T}^{(0)}e^{iW_D}
   =\frac{z_1z_2 e^2\gamma^0}{m^2+2p^2(1-\cos\theta)}e^{iW_D}\, ,
\end{equation}
where the \emph{Dalitz phase} is, after generalizing $p/E\to\beta_{12}$ and including an $\epsilon=0^+$ regulator as in \cref{eq:phase-coef-W},
\begin{equation}
\label{eq:DalitzPhase}
iW_D=\Big(\epsilon-i\frac{z_1z_2\alpha }{\beta}\Big)
\ln\bigg[\frac{2p^2}{m^2}(1-\cos\theta)\bigg]
\end{equation}
and recognizes this would reproduce the non-relativistic exact solution.\footnote{Dalitz observes that the additional phase factor,  $\exp(iW_D)$, ``bares some similarity to the term $\ln(2pr)$  of [the non-relativistic Coulomb problem's] exact solution."
} 
This result suggests that the phase $\exp(iW)$ is present even for spin-$\frac12$ particle scattering and when including some retardation effects.

Further evidence for the presence of the phase $\exp(iW)$ can be gleaned from Weinberg's work on IR divergences due to massless mediators \cite{Weinberg:1965nx}.  He notes that the coefficient of $\ln(m)$ in Dalitz's result is consistent with his computed IR divergent phase, 
\begin{equation}
\label{eq:Weinberg}
    S_c=\prod_{k<j}\exp\bigg[-i\frac{z_k z_j\alpha}{2\beta_{kj}}\ln\Big(1+\frac{\Lambda^2}{m^2}\Big)\bigg]\,
\end{equation}
where $kj$ represents particles pairs both in the initial or both in the final state and $\beta_{kj}$ stands for their relative velocity. Moreover, Weinberg asserts that his method gives the same phase factor for any spin. This again suggests that the correction to the tree-level, $t$-channel, photon exchange amplitude is given by \cref{eq:T-corrected,eq:phase-coef-W}.
 The numerator factor in \cref{eq:phase-coef-W} should be replaced by the strength of the interaction of the exchanged mediator, as appropriate. The case of gluon exchange is more subtle and we defer it to further study. 

Let us now explore how this may be of use. Tree-level unitarity was famously used to place a limit on the mass of the Higgs boson prior to its discovery \cite{Dicus:1973gbw,Lee:1977eg,Lee:1977yc}. As long as the CM momentum $p$ is smaller than the Higgs mass $m_H$, the amplitudes for scattering of longitudinally polarized gauge bosons and/or the Higgs boson computed at tree level grow with increasing $p$. For $m_H$ that exceeds some critical value, $m_c$, and for sufficiently large $p\le m_c$ partial-wave amplitudes violate the inequality \cref{eq:uni-bound-aJ}, and this leads to the perturbative unitarity bound $m_H\le m_c$.   Unitarity is guaranteed for SM amplitudes, and it can be shown on general grounds that amplitudes do not grow indefinitely and never violate the unitarity bound \cite{Cornwall:1974km,Vayonakis:1976vz}. For $m_H>m_c$ the corrections to tree-level amplitudes become significant and the model becomes non-perturbative. Therefore, for a Higgs mass that satisfies the bound $m_H<m_c$ the model remains perturbative, while for $m_H>m_c$ one obtains a strongly coupled Higgs sector.

As Ref.~\cite{Lee:1977eg}, hereafter LQT,  explains, one may characterize the growth of tree-level partial-wave amplitudes according to 
\begin{equation}
    a_\ell=A(p/m_W)^4+B(p/m_W)^2+C\,.
\end{equation}
Following LQT, we refer to these as $A$, $B$ and $C$ ``forces". Of the processes considered there, we will only consider $W_L^+W_L^-\to W_L^+W_L^-$ scattering because only for this process does a photon $t$-channel exchange contribute to the amplitude. $A$  forces cancel among graphs involving only gauge bosons while $B$ force cancellation occurs only after adding to these Higgs exchange graphs.  The determination of $m_c$ is obtained by requiring that the $C$ forces do not exceed the unitarity bound in \cref{eq:uni-bound-aJ}. However, the tree-level $t$-channel photon exchange contribution makes the partial waves ill-defined. Even worse, in the large $p$ expansion the $Z$-exchange contribution also gives a divergent contribution to the $C$ forces. LQT circumvents these difficulties by focusing its attention on the contribution to the $C$ force proportional to $m_H^2$, and ignoring higher orders in a $1/p$ expansion that are both $m_H$ dependent and increasingly singular as $\cos\theta\to1$. 

The full expression for the tree-level scattering amplitude is a bit unwieldily, and we do not reproduce it here. A much simpler expression is obtained in the limit $p\to\infty$ at fixed $\theta\neq0$:

\begin{equation}
    \label{eq:WW-high-pCM}
\mathcal{T}^{0}=-\sqrt{2}G_F\left(  2m_H^2-  \frac{\left(3+\cos^2\theta\right)}{(1-\cos\theta) } m_Z^2
\right)\,.
\end{equation}
Retaining only the term proportional to the Higgs mass, the $\ell=0$ scattering amplitude is
\[
a^0_{\rm Higgs}=-\frac{G_Fm_H^2}{4\sqrt2\pi}\,.
\]
Applying the unitarity bound in \cref{eq:uni-tree-bound-aJ} to this expression, LQT obtains
\begin{equation}
    \label{eq:LQT-uni-bound1}
    m_H^2\le \frac{4\sqrt2\pi}{G_F}\,.
\end{equation}

Were we to attempt to compute the partial-wave amplitude from the full expression in \cref{eq:WW-high-pCM}, we would encounter a divergent angular integral. Using the prescription in \cref{eq:T-corrected,eq:phase-coef-W} we tame the singularity:
\begin{align}
 a_0
    &=\frac{-G_F}{4\sqrt{2}\pi}\int_0^1 \!\!dx\;
    e^{(\epsilon-i\alpha_2)\ln(x)}\Bigg[m_H^2
    -\left(\frac1x-1+x\right)m_Z^2\Bigg] \nonumber\\
 \label{eq:a0-WW}   
 &=\frac{G_F}{4\sqrt{2}\pi}\Bigg[i\frac{m_Z^2}{\alpha_2}
    - m_H^2-\frac12m_Z^2 + \mathcal{O}(\alpha_2) \Bigg]\,,
\end{align}
where in the phase $W$ we have used $\beta_{12}\approx1$, since $p\to\infty$. Note that in this limit the $W^3$ boson is responsible for the $t$-channel pole and thus the coupling  is $g_2$, $\alpha_2=g_2^2/4\pi$.
The integral is rendered finite by including the phase $e^{iW}$ in \cref{eq:T-corrected} and taking the limit $\epsilon\to 0^+$.

Imposing \cref{eq:uni-bound-aJ} on the partial wave in \cref{eq:a0-WW}, we obtain $m_H\le {1160}\;\text{GeV}$
to be compared with \cref{eq:LQT-uni-bound1} which gives $m_H\le\;1234\text{ GeV}$.\footnote{The best bound quoted in LQT, $m_H^2\le \frac{8\sqrt2\pi}{3G_F}\simeq 1007\;\text{GeV}$, is obtained by diagonalizing the coupled channels $W_L^+W_L^-$, $Z_LZ_L$ and $hh$. In the absence of the (ill-defined) $W_L^+W_L^-$ channel, the remaining two channels yield the bound in \cref{eq:LQT-uni-bound1}.  }  
The non-perturbative nature of the result \cref{eq:a0-WW} is now evident in the term $\propto m_Z^2/\alpha_2$. However, the result cannot be trusted. The reason is that  the large-$p$ limit neglects a series in powers of $[p^2(1-\cos\theta)]^{-1}$ which for any finite (albeit large) $p$ diverges upon integration over angles even in the presence of the phase factor $e^{iW}$. 

\begin{figure}
    \centering
    \includegraphics[width=
    \linewidth]{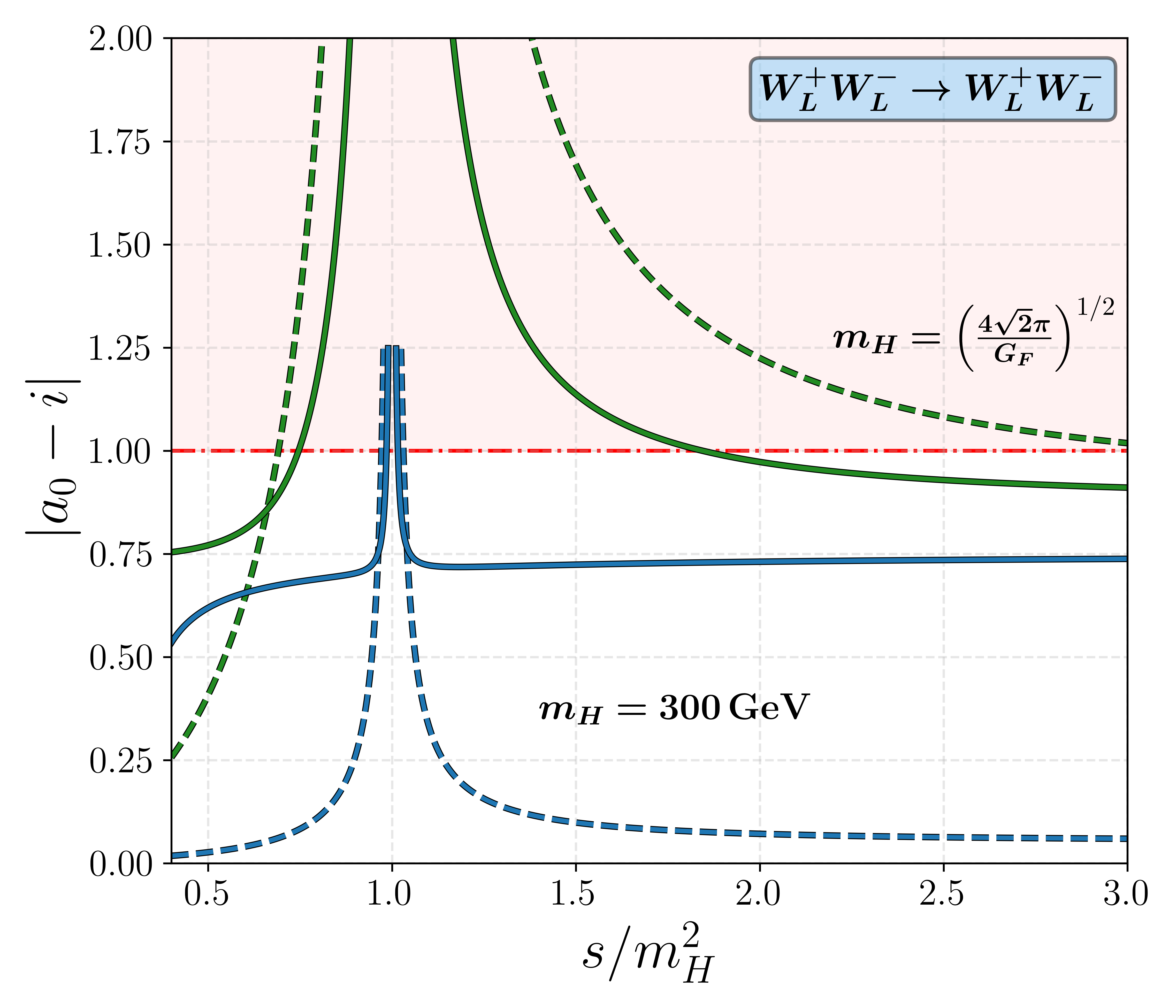}
    \caption{Comparison of the perturbative unitarity bound from Lee, Quigg and Thacker (LQT) \cite{Lee:1977eg} (dashed) with our improved quasi-perturbative unitarity bound (solid) as a function of $s/m_H^2$. Following \cite{Lee:1977eg}, we show the $W_L^+W_L^- \to W_L^+W_L^-$, $J=0$ partial-wave amplitude $a_0$ for $m_H=300$\;GeV (blue) and $m_H^2=4\sqrt2\pi/G_F$ (green). As opposed to LQT we include all contributions to the tree level scattering amplitude, and tame the forward singularity by including the effect of the multiple photon resummation in \cref{eq:T-corrected,eq:phase-coef-W}.
    {\hfill}} 
    \label{fig:LQTlike}
\end{figure}

Instead,  we are in a position to use the full expression of the scattering amplitude for arbitrary $p$. This includes, among others, the $t$-channel photon exchange that was missing in the LQT bound. The phase $e^{iW}$ has the photon coupling,  $\alpha_{\rm em}$, as it arises from multiple massless particle exchanges between the in- and out-going charged states. \cref{fig:LQTlike} shows the result of the calculation, as a function of $s/m_H^2$, for two values of the Higgs mass, $m_H$.  Except for the obvious violation of unitarity at the $s$-channel pole, which would be eliminated if the Higgs width were included, $a_0$ satisfies the unitarity bound for any momentum as long as the Higgs mass is sufficiently small. Examining the large $p$ behavior one obtains the {\it quasi-perturbative unitarity} bound
\begin{equation}
m_H\lesssim 1420\;\textrm{GeV}, 
\end{equation}
which, although weaker than LQT's, does not neglect any tree-level contributions, nor does it rely on a troubling large $p$ expansion.

It is interesting to note that an amplified contribution to the partial waves similar to that in \cref{eq:a0-WW} can be present in resonant scattering in the $s$-channel. Consider a pure gauge version of the SM with $m_Z>2M_W$, \ie\, $\cos\theta_W\leq1/2$, so that on-resonance $WW$ scattering in the $s$-channel is accessible.  The denominator of the $Z$-propagator is  given by the imaginary part of the on-shell self-energy, which we calculate at 1-loop. The resulting perturbative unitarity constraint excludes the parameter region $\cos\theta_W\le1/2$, which coincides with the kinematic condition above. Hence, for $m_Z>2m_W$ non-perturbative effects must be included in the calculation of the scattering amplitude in order to preserve unitarity. In this case,  of course, there is a good non-perturbative candidate for unitarity restoration:  stable $W^+$-$W^-$ pairs form bound states of masses $\le 2m_W<m_Z$. 

Our methods can be used as well for $t$-channel exchange of massive particles, when their mass is much smaller than the CM momentum, $m\ll p$. This can be of use in the calculation of scattering amplitudes mediated by $t$-channel exchange of Weakly Interacting Slim Particles (WISPs) such as dark photons~\cite{Holdom:1985ag} and Axions or Axion-Like Particles (ALPs)~\cite{Peccei:1977hh,Weinberg:1977ma,Wilczek:1977pj,Dine:1981rt}, among many. Absent the correction \cref{eq:T-corrected}, the resulting partial waves, albeit finite, display non-physical sensitivity on $m$. Instead, when our correction is included, the spurious sensitivity on $m$ disappears: the calculation can be performed at vanishing mass as in the examples above. Consider the scattering of a charged massive scalar of mass $M$, via a photon of mass $m$ and coupling $g$. Using the Dalitz phase in \cref{eq:DalitzResum,eq:DalitzPhase} one can readily compute 

\begin{align*}\displaybreak[1] 
    a_0&=\frac{\alpha}{8}\int_{0}^1\!\!\! dx \;\frac{4(M^2+(2-x)p^2)}{m^2+4p^2x}e^{(\epsilon+i\alpha/\beta)\ln(\frac{4p^2x}{m^2})}\\
    &=-\frac{\alpha}{8} (-1)^{\epsilon+i\alpha/\beta} \bigg[\frac{m^2}{4p^2} B\Big({-\frac{4 p^2}{m^2}};2+\epsilon+ \frac{i\alpha}{\beta },0\Big)\nn\\
    &\qquad\qquad + \frac{M^2+2 p^2}{p^2} B\bigg({-\frac{4 p^2}{m^2}};1+\epsilon+ \frac{i\alpha}{\beta },0\bigg)\bigg]\\
 &= i \frac{2M^2+p^2}{16p^2}\left[\beta-
     \left(\frac{4p^2}{m^2}\right)^{\!\!\epsilon+i\alpha/\beta} 
   \!\! +\mathcal{O}\left(\alpha,\frac{m^2}{p^2}\right)\right]
    \end{align*}
where $\beta=p\sqrt{p^2+m^2}/(p^2+m^2/2)$, $\alpha\equiv g^2/4\pi$ and $\mathrm{B}(x ; a, b)$ is the incomplete Euler Beta function. The order of limits is crucial: as $m \to 0$ for fixed $\epsilon$ we recover the massless photon quasi-perturbative amplitude, while if $\epsilon\to0$ first the amplitude displays the expected logarithmic sensitivity, $\sim \ln(m)$. This is not surprising: already in the non-relativistic problem, the long range Coulomb potential does not admit an asymptotically plane wave solution, while the short range Yukawa potential does, regardless of how small the non-vanishing ``photon''  mass may be. 

In this work, we have proposed a way of dealing with the divergences associated with massless propagators in the $t$-channel in the computation of partial-wave amplitudes. By incorporating an oscillating phase, we are able to tame the divergence, rendering it integrable. Still a first-principle derivation of this method remains to be demonstrated.
Applying this method to longitudinal $W^+W^-$ elastic scattering, we derive an improved quasi-perturbative bound on the Higgs mass to be compared to the classical result of LQT \cite{Lee:1977eg}. 
The results of this work have wider applications ranging from quasi-perturbative unitarity bounds on EFTs including massless particles to extensions of the SM containing very light massive particles.

\section*{Acknowledgements}
We warmly thank Xiaochuan Lu for his involvement in the early stages of the project and for the insightful contributions he provided, and Javi Serra y Gregorio Herdoíza for conversations. The work of B.G. and P.Q. is supported by the U.S. Department of Energy under grant number DE-SC0009919. The work of M.F.Z. is supported by the Spanish MIU through the National Program FPU (grant number FPU22/\hspace{0pt}03625) and by the Spanish Research Agency (Agencia Estatal de Investigaci\'on) through the grant IFT Centro de Excelencia Severo Ochoa No CEX2020-001007-S and by the grant PID2022-137127NB-I00 funded by MCIN/\hspace{0pt}AEI/\hspace{0pt}10.13039/501100011033. B.G., M.F.Z. and P.Q. acknowledge partial support by the European Union's Horizon 2020 research and innovation programme under the Marie Sk\l odowska-Curie grant agreement No.~101086085-ASYMMETRY and under the Marie Sk\l odowska-Curie Postdoctoral Fellowship grant agreement No 101207780 - AxionCount (only P.Q.). M.F.Z. thanks the Department of Physics of the University of California San Diego for hospitality during the development of this project.

\bibliographystyle{apsrev}
\bibliography{biblio}

\end{document}